\documentclass[preprint,amsmath,amssymb,prl,showpacs]{revtex4}

\usepackage[english]{babel}
\usepackage[applemac]{inputenc}
\usepackage[T1]{fontenc}
\usepackage{amsmath}
\usepackage{verbatim}
\usepackage[pdftex]{graphicx}
\usepackage[normalem]{ulem}
\usepackage{color}

\definecolor{dgreen}{rgb}{0.0, 0.7, 0.0}

\begin{document}

\bibliographystyle{aip}

\title{Power dependence of Klyshko's Stokes-anti-Stokes correlation in the inelastic scattering of light}

\author{Carlos A. Parra-Murillo, Marcelo F. Santos, Carlos H. Monken, Ado Jorio$^*$}

\date{\today}

\affiliation{Departamento de F\'isica, Universidade Federal de Minas Gerais, Belo Horizonte, MG, 31270-901, Brazil\\
$^{*}$Correspondence should be addressed to \textbf{adojorio@fisica.ufmg.br} \\
}

\begin{abstract}
The Stokes and anti-Stokes components in the inelastic scattering of light are related to phonon statistics and have been broadly used to measure temperature and phonon lifetimes in different materials. However, correlation between the components are expected to change the Stokes/anti-Stokes intensity ratio, imposing corrections to the broadly used Bose-Einstein statistics. Here the excitation power dependence of these scattering processes is theoretically described by an effective Hamiltonian that includes correlation between the Stokes and the anti-Stokes events. The model is used to fit available experimental results in three-dimensional diamond and two-dimensional graphene, showing that the phenomenon can significantly increase in the low-dimensional system under specific resonance conditions. By setting the scientific basis for the Stokes-anti-Stokes correlated phenomenon, the use of the Bose-Einstein population function for reasoning the inelastic scattering is generalized, providing a model to predict the conversion of optical phonons into heat or light, according to coupling constants and decay rates. The model applies to inelastic scattering in general.



\end{abstract}


\maketitle

The inelastic scattering of light by matter \cite{raman} exhibits two components: the {\it Stokes} ({\it S}), where an incident photon is converted into a phonon and a red-shifted {\it S} photon, and the {\it anti-Stokes} ({\it aS}), where one incident photon and one existing phonon are annihilated, generating a blue-shifted {\it aS} photon. The Stokes/anti-Stokes intensity ratio is a signature of the quantum character of lattice vibrations, considered to be defined by the Bose-Einstein distribution
\begin{equation}
\frac{I_{aS}}{I_{S}} = C\frac{n_0}{1+n_0},
\label{eq-BoseEinstein}
\end{equation}
where $C$ depends on the optical setup, and $n_0 = (e^\frac{\hbar \nu}{k_BT}-1)^{-1}$ is the effective phonon population \cite{raman-book}. $n_0 \ll 1$ for phonons with energies $(\hbar \nu)$ larger than the thermal energy $k_BT$ ($k_B$ is the Boltzmann constant). The $I_{aS}/I_{S}$ intensity ratio is used to measure phonon lifetimes \cite{song2008}, local effective temperatures \cite{faugeras2010,berciaud2010} and optical resonances \cite{agsf2001}, and except for very specific resonance conditions, the $I_{aS}/I_{S}$ ratio should approach the unit only if the temperature is high enough to activate a very large phonon population. However, Klyshko \cite{Klyshko-1,klyshko} has proposed a correlated process, called here the {\it Stokes-anti-Stokes} ({\it SaS}) event, where a phonon created by the Stokes process is subsequently annihilated in the anti-Stokes process. If the {\it SaS} event is significant, the picture described by the Bose-Einstein phonon distribution is not complete, and Eq.\,\ref{eq-BoseEinstein} has to be generalized.

Evidences for the {\it SaS} process are accumulating in materials science \cite{song2008,berciaud2010,Lee2012,jorio2014,kasperczyk2015}, generating interests in quantum optics \cite{Lee2012,kasperczyk2015}. Lee {\it et al.} \cite{Lee2012} demonstrated the quantum nature of the {\it SaS} correlation in diamond by measuring a non-classical {\it SaS} field correlation function $g^{(2)}$. Klyshko \cite{klyshko} pointed that the correlated character of this {\it S} and {\it aS} photons can be continuously varied from purely quantum to purely classical and Kasperczyk {\it et al.} \cite{kasperczyk2015} explored this transition in character by changing the photon and phonon reservoirs through changes in the excitation laser power. While these experiments are usually performed with ultra-fast pulsed lasers to enhance the response of the non-linear {\it SaS} event, Jorio {\it et al.} \cite{jorio2014} have provided evidences for the observation of dominant {\it SaS} event using a few miliwatts continuum wave (CW) laser, i.e. achievable even with a simple laser pointer. This result was obtained in twisted-bilayer graphene (tBLG), a two-dimensional system specially engineered to exhibit resonance with the {\it aS} photon emission \cite{jorio2014}. Since phonons have a significant lifetime, it has been proposed that these systems can work as a solid state quantum memory, storing information between the write ({\it Stokes}) and read ({\it anti-Stokes}) processes \cite{Lee2012}. In diamond and graphene $\hbar \nu \gg k_BT$, and the quantum memory would be able to work at room temperature.


Here the $I_{aS}/I_{S}$-based phonon population analysis is generalized by proposing an effective Hamiltonian that explicitly considers the Stokes and anti-Stokes fields correlation, on a simple formalism that is able to fit the experimental results from diamond and graphene. The Hamiltonian is given by
\begin{eqnarray}\label{eq:01}
 \hat H  & =&  \hbar \omega_0 \hat a^{\dagger}\hat a + \hbar \nu \hat c^{\dagger}\hat c + \hbar \omega_{S} \hat b_S^{\dagger}\hat b_S
 + \hbar \omega_{aS} \hat b_{aS}^{\dagger}\hat b_{aS}\nonumber\\
&&+ \hbar \lambda_S (\hat a \hat c^{\dagger}\hat b_{S}^{\dagger} + h.c.)+ \hbar \lambda_{aS} (\hat a \hat c \hat b_{aS}^{\dagger} + h.c.),
\end{eqnarray}

\noindent where $\hat b_{S,aS}$ ($\hat b_{S,aS}^{\dagger}$), $\hat c$ ($\hat c^{\dagger}$) and  $\hat a$ ($\hat a^{\dagger}$)
represent the annihilation (creation) operator of Stokes ({\it S}), anti-Stokes ({\it aS}), phonon ({\it c}) and incident ({\it a}) fields
respectively. $\lambda_S$ and $\lambda_{aS}$ are the coupling constants. The $\lambda_S$ coupling term describes the
creation of an {\it S} photon and a phonon through the absorption of an incident photon, and the $\lambda_{aS}$ term describes the creation of an {\it aS} photon through the absorption of an incident photon and a phonon. $\omega_0$ and $\nu$ are the incident and the phonon fields frequencies,
respectively, then the Stokes and anti-Stokes modes have energies given by $\hbar \omega_{aS,S}=\hbar \omega_0 \pm \hbar \nu$.
Typically, the coherent sources used in experiments of this type can be considered to have large enough
number of photons, which allows us to replace $(\hat a,\hat a^{\dagger})\rightarrow |\alpha|$, with $|\alpha|^2$
being the mean number of incident photons. The laser power is given by $P_L={\cal A}|\alpha|^2$, ${\cal A}$ being a constant with power units, depending on the laser frequency. The Hamiltonian model above is valid within the coherence time of the pumping laser, whether continuum or pulsed.

Hamiltonian (\ref{eq:01}) accounts for the interaction at the material but it does not properly describe the dissipation of the created excitations due to the presence of
the photonic and phononic reservoirs. Such dynamics can be computed by means of the Markovian master equation
\cite{Carmichael} for the overall reduced density operator of the three fields, introducing decay rates of phonons
and scattered photons. The master equation for the density operator $\hat\rho=\text{Tr}_R\hat\rho_{\rm total}$,
in the Lindblad form, reads
\begin{equation}\label{eq:02}
 \frac{d}{dt}\hat\rho = -i[\hat H,\hat\rho]+\mathcal{L}(\hat\rho)\,
\end{equation}
\noindent where $\mbox{Tr}_R$ indicates the tracing out of the reservoir degrees of freedom.
The Lindbladian term in our model is $\mathcal{L}=\mathcal{L}_b+\mathcal{L}_c$, with:
\begin{eqnarray}\label{eq:03}
 \mathcal{L}_b(\hat \rho)&=&-\sum_{x=S,aS}\gamma_{x}(\hat b_{x}^{\dagger}\hat b_{x}\hat\rho+\hat\rho\hat b_{x}^{\dagger}\hat b_{x}-2\hat b_{x} \hat\rho \hat b_{x}^{\dagger})\nonumber\\
 \mathcal{L}_c(\hat \rho)&=&-\gamma_c(n_0+1)(\hat c^{\dagger}\hat c\hat\rho+\hat\rho\hat c^{\dagger}\hat c-2\hat c \hat\rho \hat c^{\dagger})\nonumber\\
& &-\gamma_cn_0(\hat c\hat c^{\dagger}\hat\rho+\hat\rho\hat c\hat c^{\dagger}-2\hat c^{\dagger}\hat\rho \hat c),
\end{eqnarray}
\noindent with $\gamma_{S}$, $\gamma_{aS}$ and $\gamma_c$ being the decay rates (proportional to the inverse of the coherence time) of the respective Stokes, anti-Stokes
and phonon fields. To analyze the Stokes and anti-Stokes
field intensities and their respective correlation function at zero delay $(\tau=0)$, we compute the average values
\begin{equation}\label{eq:04}
\langle n_{S,aS}\rangle = \langle \hat b_{S,aS}^{\dagger}\hat b_{S,aS}\rangle,\;
\langle n_{c}\rangle = \langle \hat c^{\dagger}\hat c\rangle,\; g^2(0)=\frac{\langle \hat b_S^{\dagger} \hat b_{aS}^{\dagger}\hat b_{aS}\hat b_S\rangle}{\langle \hat b_S^{\dagger}\hat b_S\rangle\langle \hat b_{aS}^{\dagger}\hat b_{aS}\rangle}.
\end{equation}

\begin{figure}[ht]
\centering
\includegraphics[width=0.7\columnwidth]{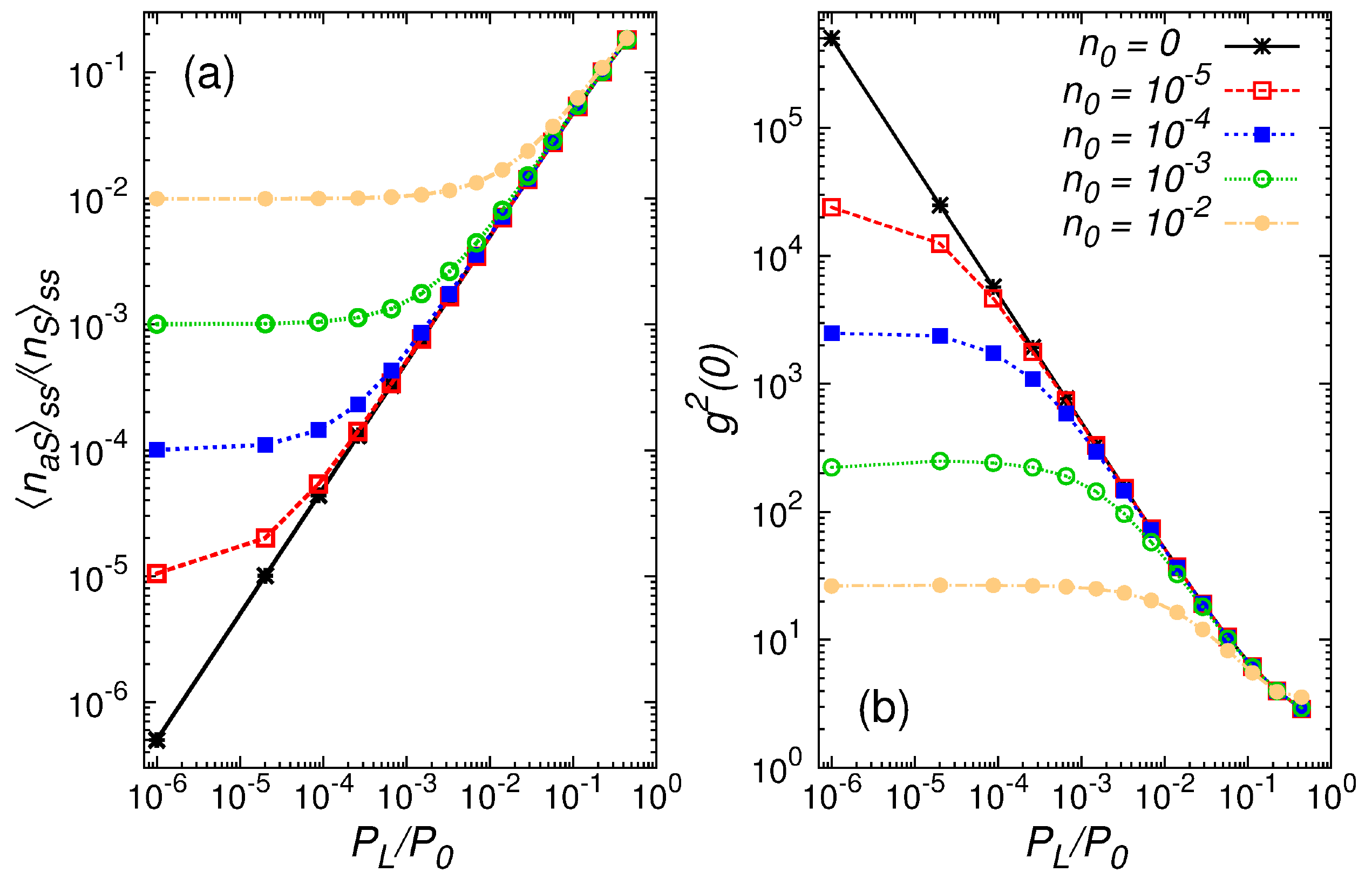}
\caption{Theoretical description of the Stokes and anti-Stokes intensity and correlation phenomena. (a) gives the population ratio $\langle n_{aS}\rangle_{ss} / \langle n_{S}\rangle_{ss}$, where $ss$ stands for steady state, and (b) gives the {\it SaS} field correlation $g^2(0)$, according with Eqs.\,\ref{eq:04}. The excitation laser power $P_L$ dependences are plot for different values of the thermal phonon population $n_0$ (see legend in (b)).
We set $\lambda_S=\lambda_{aS}=\lambda$, $\gamma_{S}=\gamma_{aS}=\gamma$
and $P_0 \approx {\cal A}\gamma\gamma_c/2\lambda^2$ for $\gamma/\gamma_c\gg n_0$. $P_0$ in power units.
The stationary phonon population gives nonphysical description for $P_L \geq P_0$ (more details in the Supplemental Information).}
\label{ExpDiamond}
\end{figure}

The power dependence for the intensity ratio $I_{aS}/I_{S}$ and the {\it SaS} field correlation $g^2(0)$ are described in panels (a) and (b) of Fig.\,\ref{ExpDiamond}, respectively, for different values of the thermal phonon population $n_0$. While the Stokes field is found to be proportional to the excitation laser power $P_L$, the anti-Stokes field exhibits two different regimes. For lower power, thermal phonons dominate the process and $\langle n_{aS}\rangle$ is proportional to $P_L$ (assuming there is no laser induced heating).
For higher powers, the {\it SaS} phenomenon dominates, and $\langle n_{aS}\rangle$ is proportional to $P_L^2$ (see Supplementary Information for details). This rationale explains the results in Fig.\,\ref{ExpDiamond}(a), $I_{aS}/I_{S}$ being constant for low power, and for high power, $I_{aS}/I_{S} \propto P_L$. In Fig.\,\ref{ExpDiamond}(b), the {\it SaS} correlation function $g^2(0)$ is shown to be proportional to the inverse laser power, and it goes to a $n_0$-dependent constant value for small values of $P_L$.

The behaviors of $I_{S}$, $I_{aS}$ and $g^2(0)$ observed by Kasperczyk {\it et al.} \cite{kasperczyk2015} are, therefore, perfectly described by the steady-state number of $S$, $aS$  photons $\langle n_{S}\rangle_{ss}$, $\langle n_{aS}\rangle_{ss}$ and $g^2(0)$ in Eqs.\,\ref{eq:04}, calculated by considering the model in Eqs.\,\ref{eq:01}-\ref{eq:04}.
Here, $\langle n_{S,aS}\rangle_{ss}$ is the steady state number of $S$,$aS$ photons.

A rationale for $I_{aS}/I_{S}$ that takes into account the {\it SaS} phenomenon can then be developed based on the theory for $\langle n_{aS}\rangle_{ss} / \langle n_{S}\rangle_{ss}$. Setting $\lambda_S=\lambda_{aS}=\lambda$, $\gamma_S=\gamma_{aS}=\gamma$, $\lambda|\alpha|/\gamma_c\neq 0$,
and considering the limit of $\gamma/\gamma_c\gg n_0$, the anti-Stokes/Stokes intensity ratio is given by  (see Supplemental Information)
\begin{equation}\label{eq:10first}
\frac{\langle n_{aS}\rangle}{\langle n_{S}\rangle} = \frac{n_0}{n_0+1} \times \frac{1+\frac{1}{2n_0}\frac{P_L}{P_0}}{1-\frac{1}{2(n_0+1)}\frac{P_L}{P_0}} \,
\end{equation}
\noindent where $P_0 \approx {\cal A}\gamma\gamma_c/2\lambda^2$ for $\gamma/\gamma_c\gg n_0$, given in power units. In the experimental range of interest, Eq.\,\ref{eq:10first} can be approximated by
\begin{equation}\label{eq:10}
\frac{I_{aS}}{I_{S}} \approx C'\frac{n_0}{n_0+1}\left[1+\left(\frac{1}{n_0}+\frac{1}{n_0+1}\right)\frac{\lambda^2}{\gamma\gamma_c}\frac{P_L}{{\cal A}}\right]\, ,
\end{equation}
\noindent where the fundamental constants $\lambda^2/{\cal A}\gamma\gamma_c \equiv C_{SaS}$ measure the importance of the {\it SaS} phenomena, and $C'=C_{aS}/C_S$, with $C_{S,aS}$ defined as the proportionality constant such that $I_{S,aS}=C_{S,aS} \langle n_{S,aS}\rangle_{ss}$. The constant $C'$ depends on the optical parameters, for instance, absorption
coefficients, inelastic scattering cross sections and the dimension (geometry) of the sample under study \cite{lo1980}.

\begin{figure}[ht]
\centering
\includegraphics[width=0.7\columnwidth]{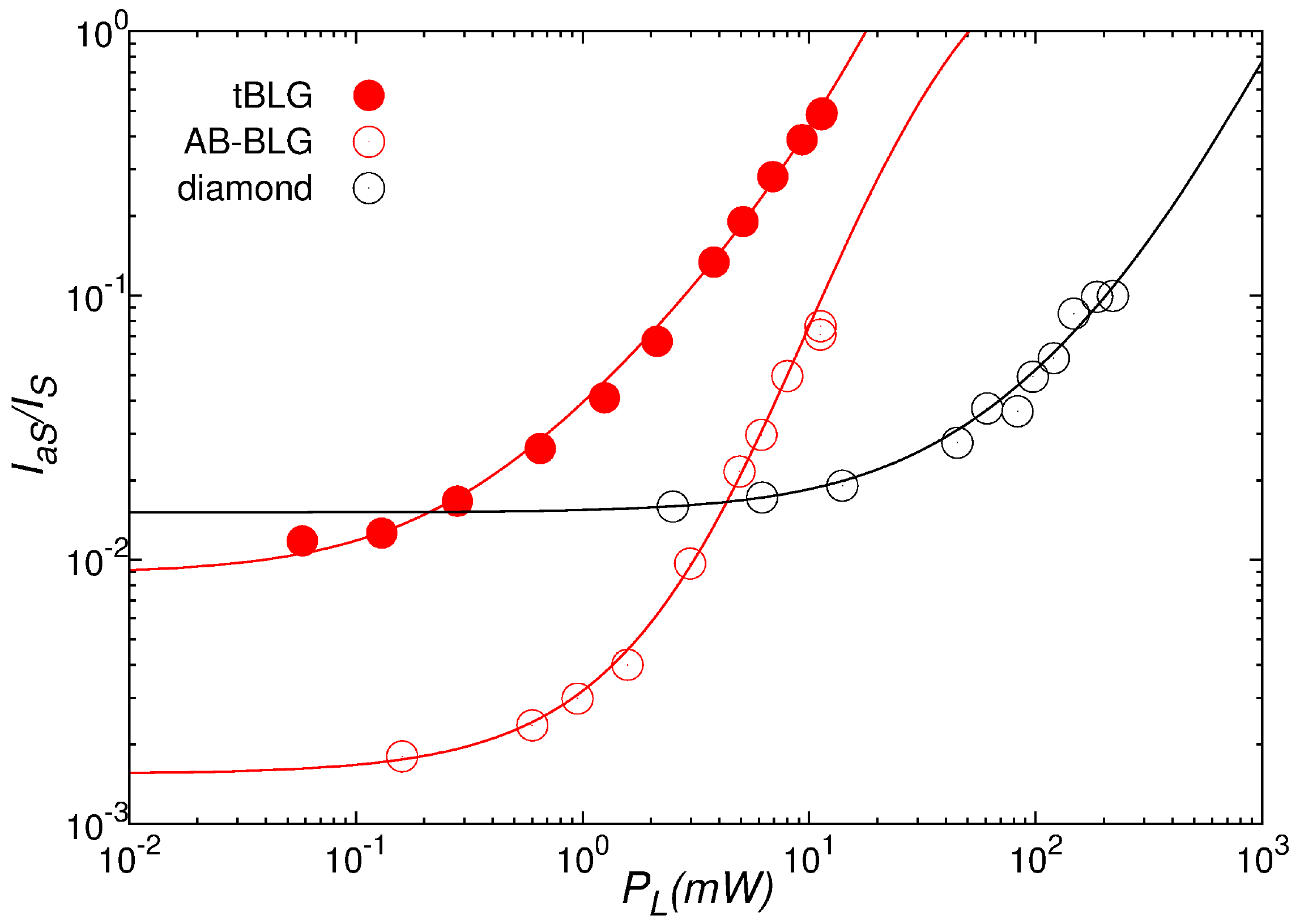}
\caption{Fitting the $I_{aS}/I_S$ intensity ratio in three different solid state systems: three-dimensional diamond (black data), two-dimensional bi-layer graphene (AB-BLG, open red data), and two-dimensional twisted bi-layer graphene (tBLG, solid red data). Data points are experimental results from Refs.\,\cite{kasperczyk2015} (diamond) and \cite{jorio2014} (graphene-systems), and lines are fit to the data using Eq.\,\ref{eq:10} and the parameters listed in table\,\ref{tab:parameters}.}
\label{ExpDiaGraph}
\end{figure}

Figure\,\ref{ExpDiaGraph} shows a comparison of the experimental $I_{aS}/I_{S}$ results from the three-dimensional diamond (black data) \cite{kasperczyk2015}, from a two-dimensional bi-layer graphene, where the layers are superposed in the so-called AB-staking configuration (AB-BLG, open red data) \cite{jorio2014}, and finally from the two-dimensional twisted bi-layer graphene (tBLG, solid red data), which was engineered to exhibit electronic resonance with the $aS$ photon emission \cite{jorio2014}.

The three different results observed for the three different samples in Fig.\,\ref{ExpDiaGraph} can all be fit with Eq.\,\ref{eq:10}, giving the parameters provided in table\,\ref{tab:parameters} (see Supplementary Information for fitting details). Analysis of the data shows that the increase in $I_{aS}/I_S$ for diamond is dominated by the {\it SaS} process ($C_{SaS} = 2.13\times 10^{-5}$\,mW$^{-1}$). Actually, the diamond data can be satisfactorily fit by imposing a constant temperature ($T = 295$\,K), with a larger {\it SaS} contribution ($C_{SaS} = 3.9\times 10^{-5}$\,mW$^{-1}$), consistent with the fact that, for measurements with a CW laser, within the same excitation power range, the $I_{aS}/I_{S}$ power dependence in diamond exhibits a constant behavior (see data in the Supplementary Information). For the AB-stacked bilayer graphene (AB-BLG), the $I_{aS}/I_S$ power dependence is dominated by heating, with a linear dependence of the effective phonon temperature on laser power ($T{\rm [K]} = 295 + 30P_L{\rm [mW]}$). Finally, the increase in $I_{aS}/I_S$ for twisted-bilayer graphene (tBLG) is dominated by the {\it SaS} process ($C_{SaS} = 1.26 \times 10^{-3}$\,mW$^{-1}$), with a contribution from laser induced heating ($T{\rm [K]} = 295 + 19P_L{\rm [mW]}$) that is consistently smaller than that from AB-BLG. Notice that the constant found to rule the temperature increase in tBLG ($19\pm 2$\,K/mW) is roughly half of the value obtained for the AB-BLG ($30.1\pm 0.2$\,K/mW), consistent with the G band frequency changes measured in tBLG and AB-BLG\,\cite{jorio2014}. Due to resonance with the {\it aS} photon emission in tBLG, the assumption $\lambda_{S}=\lambda_{aS}$ should be discussed. However, considering $\lambda_{aS} \neq \lambda_{S}$ does not change the overall picture. More details can be found in the Supplemental Information, including a comparative analysis of the thermal versus {\it SaS} contributions to $I_{aS}/I_S$ for the three samples.

Comparing the graphene samples, while in AB-BLG the generated phonons are mostly converted into heat, in tBLG they are mostly converted into light ({\it aS} photons). But the most striking result is the much larger efficiency of the {\it SaS} process in tBLG, with a $C_{SaS} = (\lambda^2/\gamma\gamma_c){\cal A}^{-1}$ value that is two orders of magnitude larger than in bulk diamond. Besides these two orders of magnitude difference, the $P_L^2$ dependence for $I_{aS}$ in diamond can only be observed with the use of femtosecond lasers, where the pulse intensity is about $10^5$ times larger than the intensity of continuous wave laser radiation of the same average power. Furthermore, in bulk diamond the number of active atoms in the focal volume is $\sim 10^7$ times larger than in the two-dimensional graphene system. Finally, the ratio $I_{aS}/I_S$ at the highest power in tBLG reaches values of $\sim$0.5, while in diamond
$I_{aS}/I_S\sim 0.1$ was reached. Putting all numbers together, correlated {\it SaS} generation per involved atom in the graphene-based system is roughly $10^{14}$ times more efficient than in bulk diamond.

Qualitatively, the reason for the striking efficiency of correlated {\it SaS} Raman scattering in the twisted bilayer graphene system might be related to a reduction in phase-space in the scattering event. In real space, the confined two-dimensional structure, which enhances electron-hole interactions in low-dimensional structures \cite{heinz-Mpoint,spataru}, leads to a large overlap between the photon and phonon wavefunctions. In reciprocal space, the logarithmically diverging two-dimensional Van-Hove singularity at the hexagonal saddle (M) point \cite{vHs} strongly enhances the generation and recombination of electron-hole pairs at the {\it aS} photon energy \cite{jorio2014,jorio2013}. The {\it SaS} event is also expected to play a role in other low dimensional materials \cite{steiner2007}, where phase-space reduction becomes important.

Raman spectroscopy is established as an important tool to study and characterize nanostructures \cite{raman-book}.
By setting the scientific basis for the {\it SaS} phenomenon, the use of the Bose-Einstein population function for reasoning the inelastic scattering is generalized, providing a model to predict the conversion of optical phonons into heat or light, according to coupling constants and decay rates.
It is surprising that the correlation between the {\it Stokes} and the {\it anti-Stokes} components can seriously affect $I_{aS}/I_{S}$, being measurable indirectly from the intensity ratio. This aspect made it possible to measure the importance of the {\it SaS} correlation in graphene, where a direct $g^2(0)$ measurement is not possible due to ultrafast luminescence \cite{lui2010,luo2012}.

The stationary phonon population can then be analytically obtained from the steady state solution of Eq.(\ref{eq:02}) and, in the limit of $\gamma/\gamma_c\gg n_0$, it is given by
\begin{eqnarray}
\langle n_c\rangle_{ss} \approx n_0\left[1+(n_0^{-1}+2)\frac{\lambda^2}{\gamma\gamma_c}\frac{P_L}{\cal A} \right].
\end{eqnarray}
The phonon population has two contributions, one thermal and the other related to the {\it SaS} phenomenon, each contribution dominating at different excitation power ranges (see a plot of $\langle n_c\rangle_{ss}$ in Fig.\,1s of the Supplementary Information). The coupling constants and the decay rates governing the phenomena can themselves be dependent on temperature, through anharmonicities or electron-phonon scattering. On the other way around,
the {\it SaS} event is an important decay channel that has to be explicitly considered when using Raman spectroscopy to extract structural and transport properties related to phonon scattering in low-dimensional structures \cite{bonini2007,steiner2009,barreiro2009}.

More generally, since the {\it Stokes} and {\it anti-Stokes} components are common to inelastic scattering processes in general, the model is not limited to the Raman scattering by phonons. It is generic, for instance applying also to the scattering of He Atoms on surfaces, to the inelastic neutron scattering, etc. Experimental and theoretical work is needed to address the physics of the $\lambda_{S,aS}$, $\gamma_{S,aS}$ and $\gamma_c$ parameters when applied to different materials and scattering phenomena.

\section{Acknowledgement}
AJ acknowledges Lukas Novotny and Mark Kasperczyk for helpful discussions.
This work had financial support from CNPq (grants 460045/2014-8, 407167/2013-7, 303471/2012-3 and 307481/2013-1).

\begin{table}[ht!]
\begin{center}
\caption{\label{tab:parameters} Parameters used to fit the data from the three materials depicted in Fig.\,\ref{ExpDiaGraph}, using Eq.\,\ref{eq:10}.
$C_{SaS}$ is given in mW$^{-1}$ ($C'$, $\lambda$, $\gamma$ and $\gamma_c$ are dimensionless, while ${\cal A}$ is given in mW).
Temperature $T$ is given in K and laser power ($P_L$) in mW. The constant multiplying $P_L$ is given in K/mW.}
\begin{tabular}{lccc}
\hline
\hline
MATERIAL & $C'$ & $C_{SaS} \equiv (\lambda^2/\gamma\gamma_c){\cal A}^{-1}$\,[mW$^{-1}$] & $T$\,[K]\\
diamond  & $(10\pm  1)$  & $(2.13\pm 1.04)\times 10^{-5}$  & $295 + (0.37\pm 0.15)P_L$\\
tBLG     & $(20\pm 1)$  & $(1.26\pm 0.01)\times 10^{-3}$ & $295 + (19\pm 2)P_L$\\
AB-BLG   & $(3.5\pm 0.2)$ & $(0.24\pm 1.36)\times 10^{-5}$ & $295 + (30.1\pm 0.2)P_L$\\
\hline
\hline
\end{tabular}
\end{center}
\end{table}



\newpage
\vspace*{5cm}

\section{Supplemental Information: Power dependence of Klyshko's Stokes-anti-Stokes correlation in the inelastic scattering of light}

Here we provide more details about the theoretical model and include some results not shown in the main text. We also present a discussion about the analysis of the experimental data from the literature.

\section{More details about the model}\label{sec:03}

To analyze the Stokes and anti-Stokes populations
and their respective correlation functions at zero time delay $(\tau=0)$, we computed the average values
\begin{equation}\label{eq:001}
\langle n_{S(aS)}\rangle = \langle \hat b_{S(aS)}^{\dagger}\hat b_{S(aS)}\rangle,\;
\langle n_{c}\rangle = \langle \hat c^{\dagger}\hat c\rangle,\; g^2(0)=\frac{\langle \hat b_S^{\dagger} \hat b_{aS}^{\dagger}\hat b_{aS}\hat b_S\rangle}{\langle \hat b_S^{\dagger}\hat b_S\rangle\langle \hat b_{aS}^{\dagger}\hat b_{aS}\rangle},
\end{equation}

\noindent which can be obtained by means of the equation
\begin{equation}\label{eq:002}
 \frac{\partial}{\partial t}\langle \hat O\rangle=i\langle [\hat H,\hat O]\rangle+\text{Tr}(\hat O\mathcal{L}{\hat\rho}),
\end{equation}

\noindent yielding the following linear system of differential equations:
\begin{eqnarray}\label{eq:003}
 \partial_t\langle n_S\rangle&=&-2 \lambda_{\alpha}\, \text{Im}\{\langle \hat c\hat b_S\rangle\}-2\gamma_S\langle n_S\rangle\nonumber\\
 \partial_t\langle n_{aS}\rangle&=&-2 \lambda_{\alpha} \,\text{Im}\{\langle \hat c^{\dagger}\hat b_{aS}\rangle\}-2\gamma_{aS}\langle n_{aS}\rangle\nonumber\\
\partial_t\langle n_{c}\rangle&=&-2\lambda_{\alpha}(\text{Im}\{\langle \hat c\hat b_S\rangle\}-\text{Im}\{\langle \hat c^{\dagger}\hat b_{aS}\rangle\})\nonumber\nonumber\\&&-2\gamma_{c}(\langle n_{c}\rangle- n_0)\nonumber\\
 \partial_t\langle \hat c\hat b_S\rangle&=&-i\lambda_{\alpha} (\langle n_S\rangle+\langle n_{c}\rangle+\langle \hat b_S\hat b_{aS}\rangle)\nonumber\\
&&-(\gamma_{c}+\gamma_S+i\omega_0)\langle \hat c\hat b_S\rangle-i\lambda_{\alpha}\nonumber\\
 \partial_t\langle \hat c^{\dagger}\hat b_{aS}\rangle&=&-i\lambda_{\alpha}(\langle \hat b_S\hat b_{aS}\rangle -\langle n_c\rangle+\langle n_{aS}\rangle)\nonumber\\&&-(\gamma_{c}+\gamma_{aS}+i\omega_0)\langle \hat c^{\dagger}\hat b_{aS}\rangle\nonumber\\
\partial_t\langle \hat b_S\hat b_{aS}\rangle&=&-i\lambda_{\alpha} (\langle \hat c^{\dagger}\hat b_{aS}\rangle+\langle \hat c\hat b_{S}\rangle)\nonumber\nonumber\\
&&-(\gamma_{S}+\gamma_{aS}+i2\omega_0)\langle \hat b_S\hat b_{aS}\rangle.
\end{eqnarray}
where  $\lambda_{\alpha} = \lambda|\alpha|$. Initially we will be setting $\lambda_S=\lambda_{aS}=\lambda$ and $\gamma_S=\gamma_{aS}=\gamma$.
The dynamics will depend on how the incident field is implemented. In case of a pulsed laser, it is enough to redefine the coupling
strength as a time-dependent function $|\lambda_{\alpha}(t)|^2=\lambda^2 |\alpha|^2 f(t)$. $f(t)$ can, for instance,
be modelled as a temporal gaussian profile, such as $f(t) = \exp(-t^2/2\sigma^2)$, where the coherence time is proportional to
$\sigma$. In that case there isn't a steady state and, in order to analyze the behavior of the field population as a
function of the laser power $P_L$, we might consider measures at the excitation time, that is, the time at which
the population gets its first maximum, or the time-averaged values of the observables defined in Eq.~(\ref{eq:001})
as $\overline{\langle\hat O\rangle_t}=\lim_{\Delta t\rightarrow \infty} \int_0^{\Delta t}\langle\hat O\rangle_tdt/\Delta t$.

We first focus on the case of a continuum pumping, i.e., $\lambda_{\alpha}(t)=\lambda_{\alpha}$, on the material.
In this case, the system of differential equations~(\ref{eq:003}) has a steady state solution due to the environment-induced
relaxation. To find them, we write the system in~(\ref{eq:003}) as
a vectorial diffential equation $\partial_t\vec{x}=M \vec{x}+ \vec{b}$, where $M$ is a time-independent square matrix and
\begin{eqnarray}\label{eq:004}
\vec{x}^{T}&=&(\langle \hat n_S\rangle,\langle \hat n_{aS}\rangle,\langle \hat n_c\rangle,
\langle \hat c\hat b_S\rangle,\langle \hat c^{\dagger}\hat b_S^{\dagger}\rangle,
\langle \hat c^{\dagger}\hat b_{aS}\rangle,\langle \hat c \hat b_{aS}^{\dagger}\rangle,\langle \hat b_S \hat b_{aS}\rangle,\langle \hat b_S^{\dagger}\hat b_{aS}^{\dagger}\rangle)\\
\vec{b}^{T}&=&(0,0,2\gamma_c n_0,-i\lambda_{\alpha},i\lambda_{\alpha},0,0,0,0),
\end{eqnarray}

\noindent and set $d_t \vec{x}_{ss}=0$, where $ss$ stands for steady state. To solve this, we invert the matrix $M$, which
is only possible if $M$ is not singular, i.e. if its determinant $Det(M)$ is different from zero.
The solutions are given by $\vec{x}_{ss}=-M^{-1}\vec{b}$. By simple inspection, it is straightforwardly noticed
that the Stokes, anti-Stokes and phonon populations satisfy an equilibrium condition
\begin{equation}\label{eq:005}
\langle \hat n_c\rangle_{ss}-n_0=\frac{\gamma}{\gamma_c}(\langle \hat n_S\rangle_{ss}-\langle \hat n_{aS}\rangle_{ss}).
\end{equation}
This equation shows that if Stokes and anti-Stokes are produced at the same rate, the only phonons left are the thermal ones. On the other hand, since in general 
Stokes production is favoured over anti-Stokes (as it becomes clear in a few lines), whenever these scattering processes take place,
the distribution of phonons in the sample is not determined solely by temperature anymore. 

One of the central results in the paper is to show the behaviour of the Stokes-anti-Stokes field intensities in terms
of the power of the pumping laser. In the steady state regime, we can obtain an analytical expression for those intensities, $I_{S(aS)}$, which are proportional to
the average population of the photon modes $\langle \hat n_{S(aS)}\rangle_{ss}$. For  $S$- and $aS$-modes we find
\begin{eqnarray}\label{eq:006a}
\langle n_{S}\rangle_{ss}=\frac{2 (1 + n_0) \tilde\gamma (1 + \tilde\gamma) \tilde\lambda_{\alpha}^2 - (3 + 4 n_0 +
    2 \tilde\gamma) \tilde\lambda_{\alpha}^4}{2 (\tilde\gamma + \tilde\gamma^2 - 2 \tilde\lambda_{\alpha}^2) (\tilde\gamma (1 + \tilde\gamma) - (1 + 2 \tilde\gamma) \tilde\lambda_{\alpha}^2)}\,,
\end{eqnarray}

\begin{eqnarray}\label{eq:007b}
\langle n_{aS}\rangle_{ss}=\frac{2 n_0 \tilde\gamma (1 + \tilde\gamma) \tilde\lambda_{\alpha}^2 - (1 + 4 n_0 -
    2 \tilde\gamma) \tilde\lambda_{\alpha}^4}{2 (\tilde\gamma + \tilde\gamma^2 - 2 \tilde\lambda_{\alpha}^2) (\tilde\gamma (1 + \tilde\gamma) - (1 + 2 \tilde\gamma) \tilde\lambda_{\alpha}^2)}\,,
\end{eqnarray}

\noindent where $\tilde\gamma=\gamma/\gamma_c$ and $\tilde\lambda_{\alpha}=\lambda_{\alpha}/\gamma_c$. Note that
$\tilde\lambda_{\alpha}^2=(\tilde\lambda^2/{\mathcal A})P_L$.
The above solutions are well behaved, i.e., $Det(M)\neq 0$, and are physically
meaningful if $\langle \hat n_{S(aS)}\rangle_{ss}\geq 0$; these conditions are fulfilled if
$\tilde\lambda_{\alpha}< \sqrt{\tilde\gamma(\tilde\gamma+1)/(1+2\tilde\gamma)}$. Therefore our description
is valid below the laser power upper bound
\begin{equation}
P_0\equiv {\mathcal A}\frac{\tilde\gamma(\tilde\gamma+1)}{\tilde\lambda^2(1+2\tilde\gamma)},
\end{equation}
Note that for low laser power both $\langle n_{S(aS)}\rangle_{ss}$ increase linearly with $P_L$.
In order to easily identify different regimes for $P_L$, the intensities ratio $I_{aS}/I_{S}$ is a
good figure of merit. If  $\tilde\gamma\gg \{n_0,1\}$, a condition fulfilled in the analysed experiments,
and $\tilde\lambda_{\alpha}\neq 0$, the ratio $\langle n_{as}\rangle_{\rm ss}/\langle n_{s}\rangle_{\rm ss}$ reduces to
\begin{eqnarray}\label{eq:008}
\frac{\langle n_{as}\rangle}{\langle n_{s}\rangle} &=& \frac{n_0}{n_0+1} \times \frac{1+\frac{1}{2n_0}\frac{P_L}{P_0}}{1-\frac{1}{2(n_0+1)}\frac{P_L}{P_0}}\\
&=& \frac{n_0}{n_0+1} \times \left(1+\frac{1}{2n_0}\frac{P_L}{P_0}\right)\left(1-\frac{1}{2(n_0+1)}\frac{P_L}{P_0}\right)^{-1}\\
&\approx& \frac{n_0}{n_0+1}\left(1+\left(\frac{1}{n_0}+\frac{1}{(n_0+1)}\right)\frac{1}{2}\frac{P_L}{P_0}
 +\frac{1}{n_0(n_0+1)}\left(\frac{1}{2}\frac{P_L}{P_0}\right)^2\right)
\end{eqnarray}

\noindent where $P_0 \approx \mathcal{A} \gamma\gamma_c/2\lambda^2$. Therefore, the intensities ratio 
takes the form
\begin{eqnarray}\label{eq:009a}
  \frac{I_{aS}}{I_{S}}=C'\frac{n_0}{n_0+1}\left[1+\left(\frac{1}{n_0}+\frac{1}{(n_0+1)}\right)C_{SaS} P_L
 +\frac{C_{SaS}^2}{n_0(n_0+1)}P_L^2\right]
\end{eqnarray}
\begin{figure}[ht]
\includegraphics[width=0.8\columnwidth]{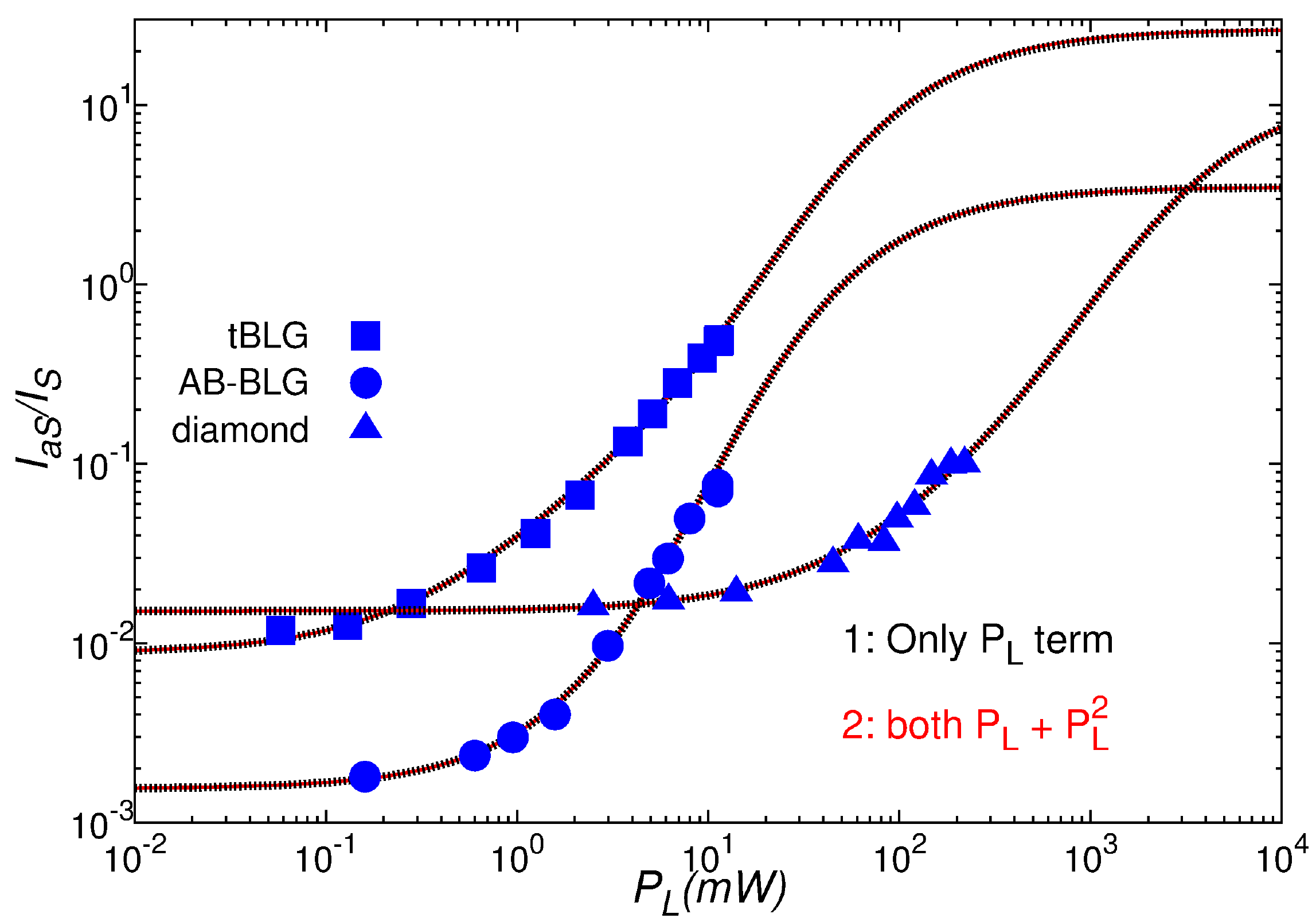}
\caption{\label{fig:01s} The figure shows the experimental measurements for the three samples investigated
in the paper (symbols). The lines are the equation (7) in paper plotted using the fitting parameter of the
table. Dashed black line corresponds to Eq. (7), red line to Eq.\,(\ref{eq:009a}) only considering the quadratic
part. Light-blue line correspond to the full equation Eq.\,(\ref{eq:009a}).}
\end{figure}

\noindent with $C_{SaS}=1/2P_0=\lambda^2/\mathcal{A} \gamma\gamma_c$. We chose to not use this notation in the paper to 
keep the equations related to the fundamental constants. Here $C'=C_{aS}/C_S$, with $C_{S(aS)}$ defined as the proportionality 
constant such that $I_{S(aS)}=C_{S(aS)} \langle \hat n_{S(aS)}\rangle_{ss}$.

In figure \ref{fig:01s} we show that the quadratic term in Eq.\,(\ref{eq:009a}) does not introduce strong corrections. In fact, all the 
experiments are done in the limit $P_L / P_0 \ll 1$ so it is reasonable that the linear term in this parameter dominates.
The blue-dashed lines corresponds to the full equation \,(\ref{eq:009a}) and the black-dashed line only to the linear
part of it, both using the fitting parameters of the table presented at the end of the paper. It is clearly seen that these both
 lines basically coincide, further justifying the approximation taken in Eq. (7) of the manuscript. 

The phonon population can also be analytically obtained from the stationary analysis and it is given by
\begin{eqnarray}\label{eq:009b}
\langle\hat n_c\rangle_{ss}=\frac{n_0 \tilde\gamma (1 + \tilde\gamma) +
(\tilde\gamma-n_0) (\tilde\lambda^2/{\mathcal A})P_L}{\tilde\gamma (1 + \tilde\gamma) -
(1 + 2 \tilde\gamma) (\tilde\lambda^2/{\mathcal A})P_L}\approx n_0 \frac{1+\frac{1}{2n_0}\frac{P_L}{P_0}}{1-\frac{P_L}{P_0}},
\end{eqnarray}

\noindent which is only valid if $\tilde\lambda_{\alpha}< \sqrt{\tilde\gamma(\tilde\gamma+1)/(1+2\tilde\gamma)}$.
In the case $\tilde\gamma \gg 1$, the phonon population reduces to
\begin{eqnarray}\label{eq:010}
 \langle n_c\rangle &\approx&n_0 \left(1+(\frac{1}{n_0}+2)C_{Sas}P_L+\frac{2}{n_0}C_{Sas}^2P_L^2\right)\,
\end{eqnarray}
Note that if $P_L\neq 0$, which activates the {\it S-aS} modes, the phonon population is always larger than $n_0$,
therefore the difference $\langle \hat n_S\rangle_{ss}-\langle \hat n_{aS}\rangle_{ss}>0$.

There are additional effects that allow us to sophisticate our effective description. For example, we can also
consider effects of phase noise caused by elastic scattering of generated phonons with different momenta. This is done
by adding an extra Lindbladian term
\begin{equation}\label{eq:011a}
\mathcal L'_c(\hat \rho) =-\gamma'_c(\hat n^2_c\hat\rho+\hat\rho\hat n^2_c-2\hat n_c\hat\rho \hat n_c).
\end{equation}

\noindent It is easily proven, using Eq.~(\ref{eq:002}), that phase noise does not modify the equations for the populations
in Eqs.~\ref{eq:006a} and \ref{eq:007b}, therefore, the equilibrium condition is preserved. Its main changes are for equations of
the expected values of the type $\langle \hat c\hat b_S\rangle$ and $\langle \hat c^{\dagger}\hat b_{aS}\rangle$,
introducing expectation values of second-order operators, as for instance $\langle \hat c\hat b_S \hat n_c\rangle$.
The new system of equations is not trivial and analytical solutions are no longer easy to obtain. In that case,
and for the computation of correlation functions $\langle \hat b_x^{\dagger} \hat b_{x'}^{\dagger}\hat b_{x''}\hat b_{x'''}\rangle$ ($x = S, aS$),
we compute the stationary solution for the density operator $\hat\rho_{ss}$, for which $d_t\hat\rho_{ss}=0$.
We must thus solve $\mathcal L_{\rm total} \rho_{ss}=0$, which requires an expansion in the Fock basis
$\{|n_S,n_{aS},n_c\rangle\}\in{\mathcal H}={\mathcal H}^{S}\otimes{\mathcal H}^{aS}\otimes{\mathcal H}^{c}$.
This basis increases exponentially and must be truncated for practical implementations. This is justifiable by the fact that since $P_L / P_0$ is always much smaller than one,
the steady state photonic population is never large. After numerically obtaining $\hat\rho_{ss}$ we can compute expected values as $\langle \hat O\rangle_{ss}={\text Tr}(\hat O \hat \rho_{ss})$,
and we are able to analyze changes on the population of the photonic modes and S-aS correlation functions in terms of the system parameters.

\section{More Theoretical Results}\label{sec:03a}

\begin{figure}[ht]
\includegraphics[width=0.85\columnwidth]{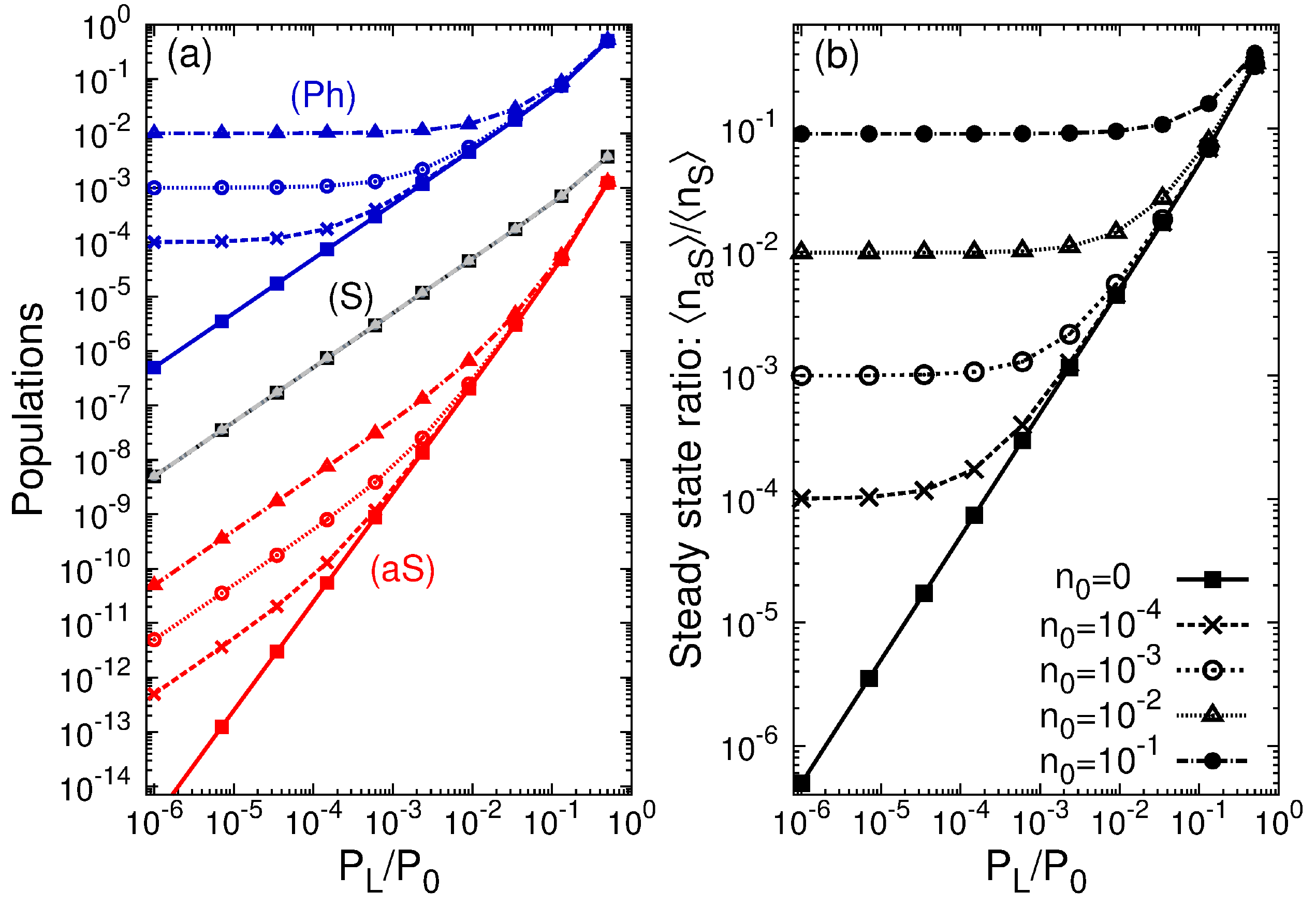}
\caption{\label{fig:02s} Panel (a) presents Stokes $(S)$, anti-Stokes $(aS)$ and phonon
$(Ph)$ populations as a function of the laser power $P_L$ in units of $P_0$. The different curves correspond
to differents temperatures set by the thermal average number $n_0$ (see legend in panel (b)). In panel (b)
we show the ratio between {\it aS} and {\it S} mode population as a function of $P_L$ and $n_0$. For this figure we set
$\tilde\gamma=100$, $\gamma'_c=0$ and a continuum pumping.}
\end{figure}

Figure~\ref{fig:02s} presents how populations depend on the laser power $P_L$ for $\tilde\gamma=100$ and different thermal
phonon numbers $n_0$. For $n_0\neq 0$ and low laser power, both photonic modes increase
linearly with $P_L$, with a slope proportional to $(n_0+1)$ for $S$-modes and
$n_0$ for $aS$-modes. In the limit of extremely low temperature ($n_0\rightarrow 0$) the aS-modes
always increase with $P_L^2$.

But for a non-zero temperature there are two regimes for the growth
of the $aS$ population, i.e.,  $I_{aS}\propto P_L$ for low laser power and $I_{aS}\propto P_L^2$
for high laser power. Note that for
$P_L\neq 0$ there exists activation of $S$- and $aS$-modes due to the inelastic scattering,
however random generation of photonic modes is also expected due to the presence of the thermal
bath of phonons. The interplay between these two processes will affect the production of photon
pair $S$-$aS$, i.e. the $SaS$ event.

\begin{figure}[ht]
\includegraphics[width=0.8\columnwidth]{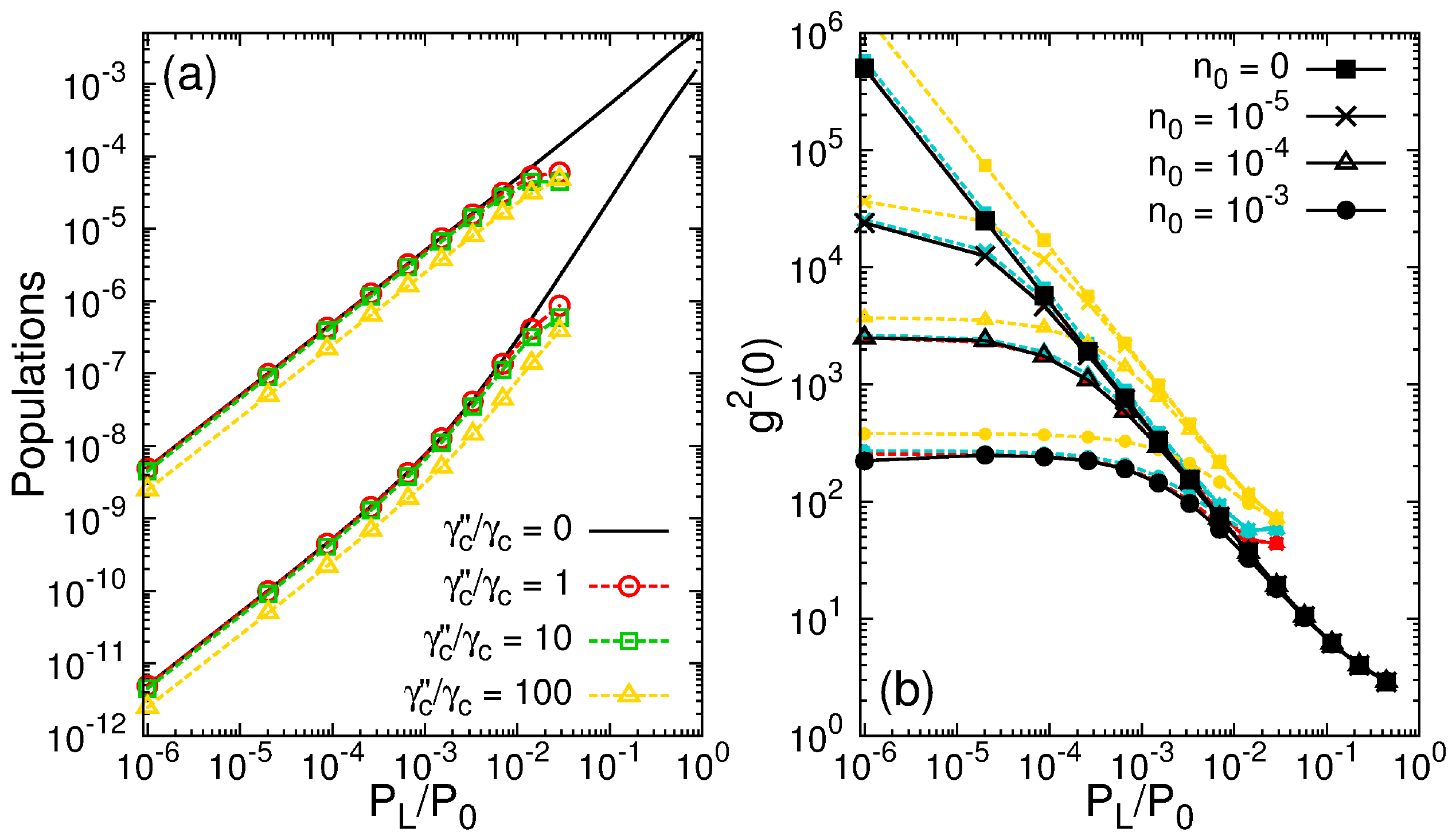}
\caption{\label{fig:03s} (a) Stokes (upper lines) and anti-Stokes (lower lines) modes
populations as a function of the normalized laser power ($P_L/P_0$) for different values of $\gamma_c'/\gamma_c$ (different
line colors, see legend). (b) The effect of $\gamma_c'/\gamma_c$ on $g^2(0)$ for different values of $n_0$ (see legends).
Other parameters are same as those in figure~\ref{fig:02s}.}
\end{figure}

The two regimes are also reflected in the results for cross-correlation function $g^2(0)$, as shown in
Fig.~\ref{fig:03s}. Note that the presence of phase noise ($\gamma_c'\neq 0$) does not affect relevantly the
phenomenology presented in this work (see Fig.~\ref{fig:03s}). Nevertheless, for very large values of
$P_L$ the convergence of our numerical fails due to the truncation of the Hilbert space, and the results are
not conclusive due to the lack of sufficient resources for tackling that regime.

\begin{figure}[ht]
\centering
\includegraphics[width=0.8\columnwidth]{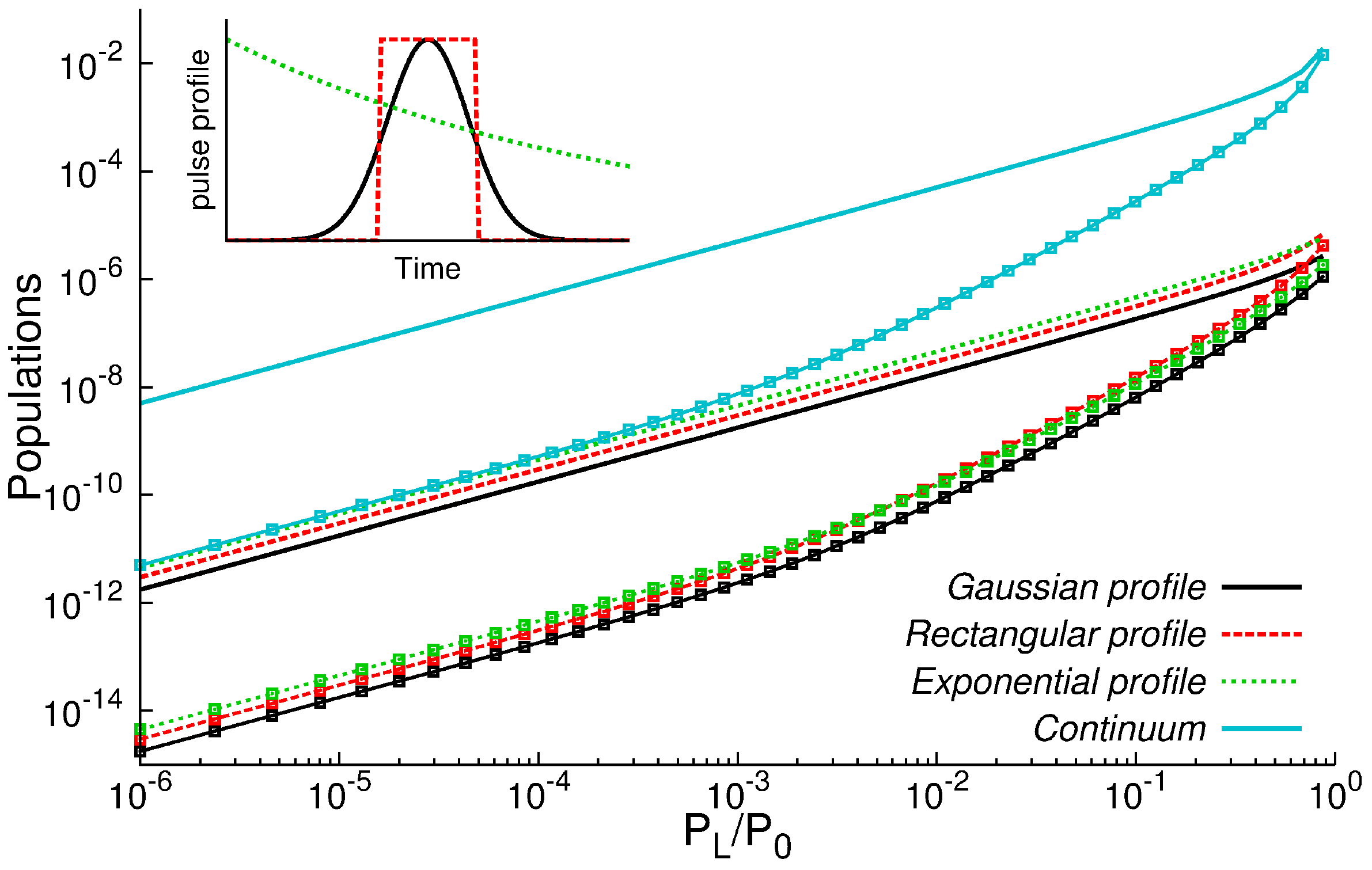}
\caption{\label{fig:04s} Stokes (lines) and anti-Stokes (lines with squares) modes
populations as a function of the normalized laser power ($P_L/P_0$) for different types of incident laser (see legend). The inset shows
the time profile of the implemented lasers. The parameters are same as those in figure~\ref{fig:02s}, with $n_0=10^{-3}$.}
\end{figure}

So far our analysis has been done for continuous pumping, which allows for well behaved steady state solutions.
When considering a time-dependent but short-time pumping, the system response is similar in shape to the
implemented pulse and an steady state is not expected. In that case, as mentioned before, we computed the
long-time averages of the observables of interest. The results for the photonic populations are shown in
Fig.~\ref{fig:04s} in comparison with the steady state solution for continuum pumping.

It is clearly seen that the phenomenology studied throughout this paper is preserved, that is, the different regimes
of the populations with the increasing laser power $P_L$ is preserved, and only changes in the intensity
of the $S$- and $aS$-modes are observed. The results in Fig.~\ref{fig:04s} allow us to generalize the findings
presented along this work.

We finally want to remark that our theory is
based on phenomenological constants that will depend on the material under study. The validity of the approximations
made to obtain analytical expressions for the ratio $I_{aS}/I_S$ and the phonon population $\langle n_c\rangle$ is
supported by typical and relevant parameters as $n_0$, $\gamma$ and $\gamma_c$. In typical experiments,
like the one reported in Ref.~\cite{Jorio-1}, $n_0=1.6 \times 10^{-3}$ at $T=295$\,K, $1/\gamma_c\approx 10^{-12}$s (phonon lifetime)
and $1/\gamma\approx 130\times 10^{-15}$s (laser pulse width).
The $\gamma/\gamma_c$ is at least one order of magnitude larger than one, and four orders of magnitud
larger that $n_0$. The incident field has a power range from 1\,mW to 1\,W and wavelength $785$\,nm. The number
of incident photons is of the order of $|\alpha|^2\sim 52\times 10^{6} - 52\times 10^{9}$ per pulse.

\section{About the Analysis of the Experimental Data} \label{sec:04}

A fair question one can ask is wether the experimental results for $I_{aS}/I_{S}$ from diamond \cite{Jorio-1} and graphene \cite{Jorio-2}
cannot be satisfactorily fit considering just the Bose-Einstein distribution function. In other words, does the data on the power dependence of
$I_{aS}/I_{S}$ from diamond and graphene provide convincing case that anything beyond laser heating has been observed?

The presence of the {\it SaS} correlation in diamond is unquestionable, given the results of the correlation function $g^2(0)$ \cite{Jorio-1}.
As discussed in our paper, our theory provides perfect description for the excitation power dependence of $I_{aS}$, $I_{S}$, $I_{aS}/I_{S}$
and $g^2(0)$. The AB-BLG is fit in our work considering that the $I_{aS}/I_{S}$ is ruled by thermal effects and, although relatively good fitting can be
obtained with a small contribution from the {\it SaS} phenomenon, we raise no controversy in this case. We now discuss results for
the tBLG sample \cite{Jorio-2}, which according to our analysis, is dominated by the {\it SaS} phenomenon.

\begin{figure}[ht]
\centering
\includegraphics[width=0.8\columnwidth]{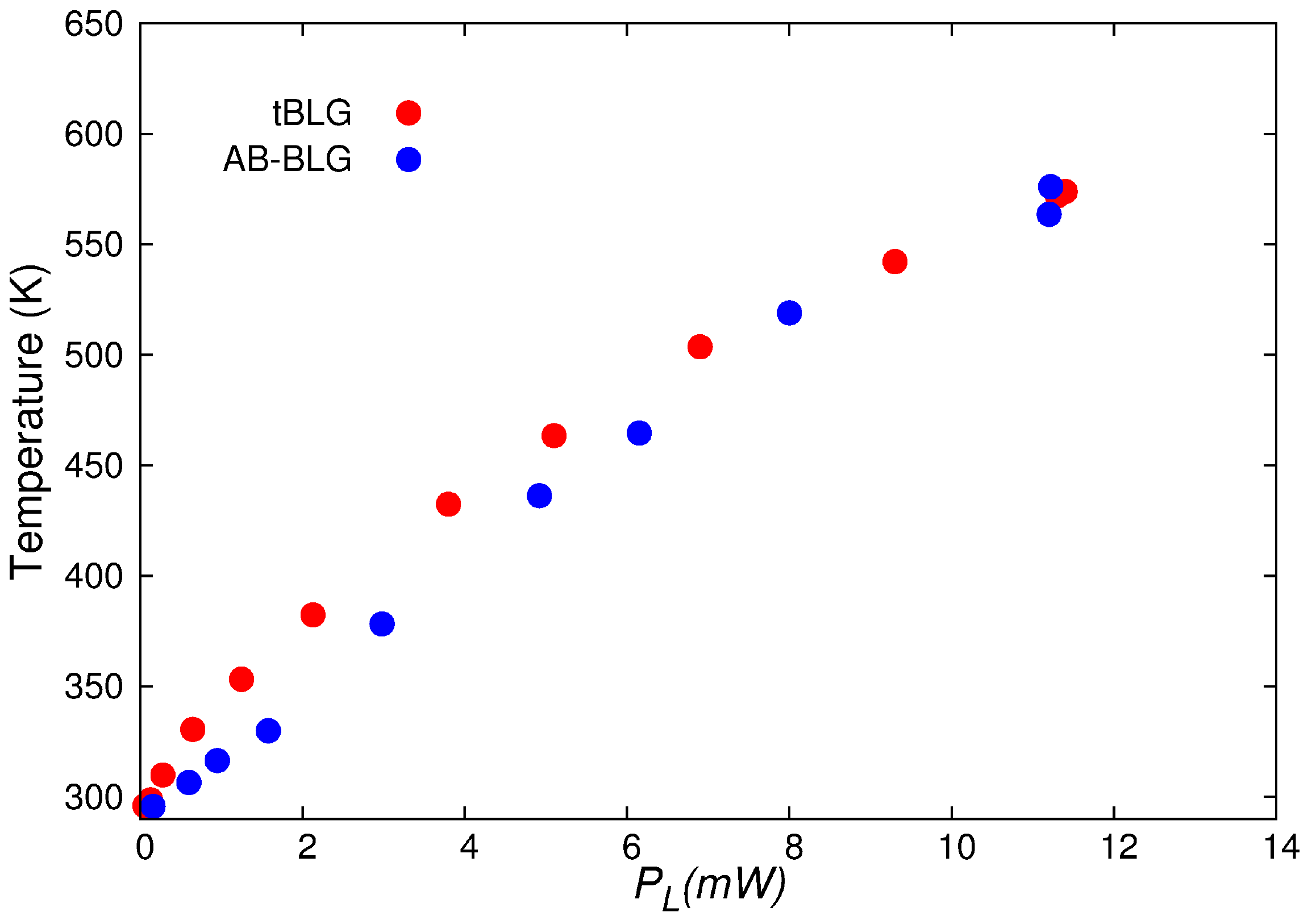}
\caption{\label{BLGallT} Expected effective temperature $T$ for tBLG (red) and AB-BLG (blue),
extracted directly from the $I_{aS}/I_{S}$ data \cite{Jorio-2}, according to Eq.\ref{eq:009a}
 neglecting the {\it SaS} phenomena.}
\end{figure}

Figure\,\ref{BLGallT} is a plot of what would be the sample temperature $T$, if extracted directly from the $I_{aS}/I_{S}$ experimental
data and using the Bose-Einstein phonon distribution function, which from Eq.(1) in the paper is given by:
\begin{equation}\label{eq:015}
T = \frac{E_{ph}}{
k_B\left[
\ln{C} -
\ln{\frac{I_{aS}}{I_{S}}}
\right]
},
\end{equation}
where $E_{ph}$ is the phonon energy, $k_B$ is the Boltzmann constant, and $C$ is the proportionality constant that accounts for
all the optical properties of the setup and sample. $C$ is chosen such that $T \rightarrow 295$\,K as $P_L \rightarrow 0$. For tBLG we obtain
$C^{tBLG} = 26$ and for the AB-BLG we obtain $C^{AB-BLG} = 4$. The fact that $C^{tBLG} > C^{AB-BLG}$ is expected, since tBLG was engineered for
specific resonance behavior. However, the fact that $T$ is generally larger for tBLG, as compared to AB-BLG (see data in Fig.\,\ref{BLGallT})
is not expected. First, the tBLG was engineered for resonance with the anti-Stokes photon emission.
Therefore it is expected that this sample would have a stronger cooling channel, and it would
actually exhibit lower temperatures than the AB-BLG. Second, this expectation is proven to be correct by the lower shift in phonon
frequency observed for tBLG, as shown in Ref.\cite{Jorio-2}. Therefore, the tBLG {\bf can not} be hotter than the AB-BLG,
and the result obtained with Eq.\,\ref{eq:15} is proven to be inconsistent.
\begin{figure}[ht]
\centering
\includegraphics[width=0.8\columnwidth]{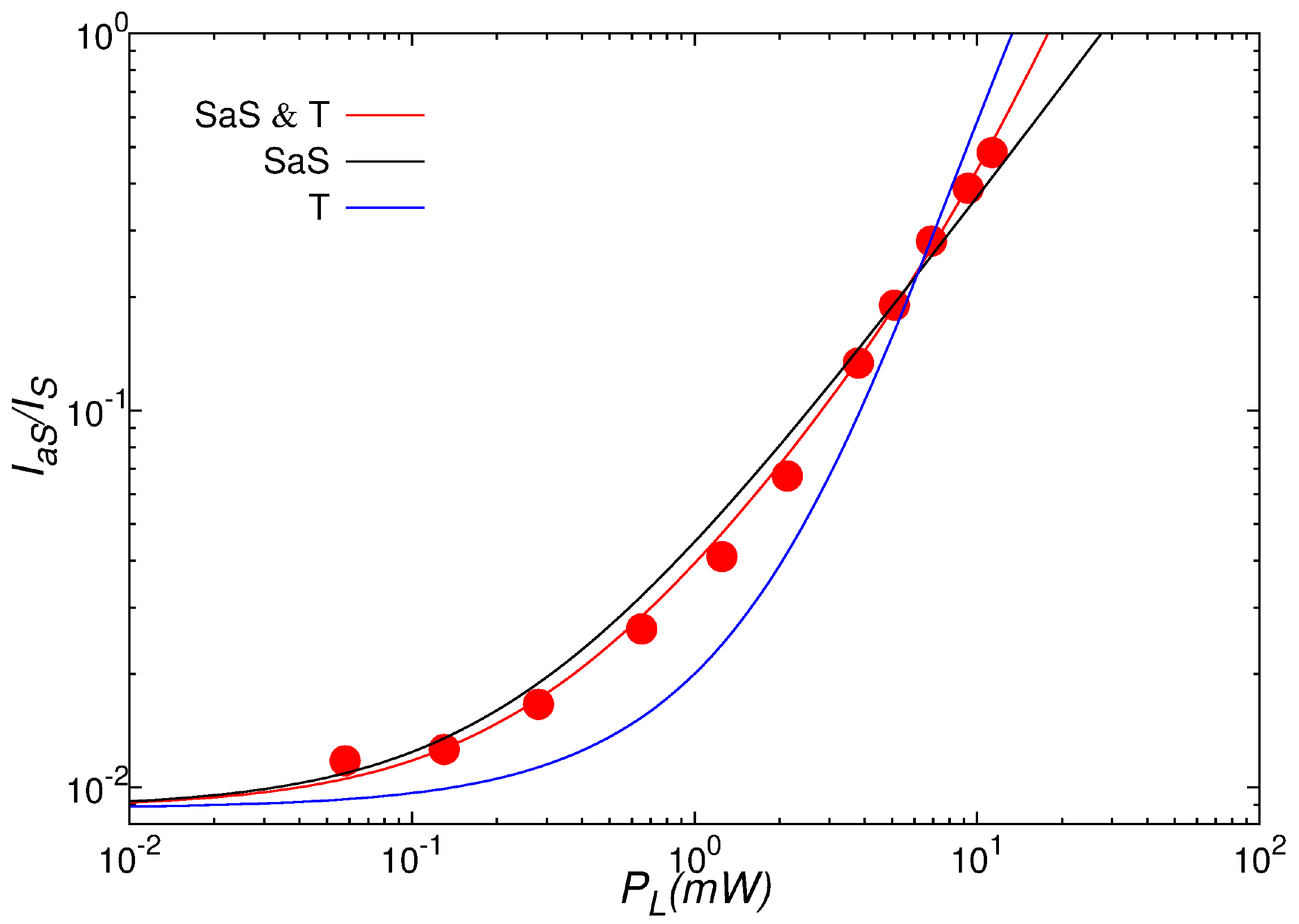}
\caption{\label{tBLG-allfits} Three different fittings for the $I_{aS}/I_{S}$ data from tBLG \cite{Jorio-2}. The red points are the
$I_{aS}/I_{S}$ data. The red line is the same fitting as in the paper. The black line is a fit to the data using Eq.(7) from the paper, with
$C'=20$, $(\lambda^2/\gamma\gamma_c){\cal A}^{-1}=1.8\times 10^{-3}$\,mW$^{-1}$ and $T=295$\,K.
The blue line is a fit to the data using Eq.(7) from the paper, with
$C'=20$, $(\lambda^2/\gamma\gamma_c){\cal A}^{-1}=0$ and $T=(295 + 35P_L)$\,K.}
\end{figure}

One can also argue on how sensitive the data fitting is with respect to the parameters choice. Parameter values outside the errors given in table I of the paper do not fit the data properly. Figure\,\ref{tBLG-allfits} shows three possible fittings of the tBLG data. The red line is the best fit, as given in the paper. The black and blue lines were obtained by considering that the $I_{aS}/I_{S}$ is ruled by pure {\it SaS} phenomena and by pure thermal phenomena, respectively. Notice that these two fitting options do not fit the data properly. The worse fitting is obtained using the pure thermal phenomena, demonstrating the predominance of the {\it SaS} phenomena.

\begin{figure}[hb]
\centering
\includegraphics[width=1.0\columnwidth]{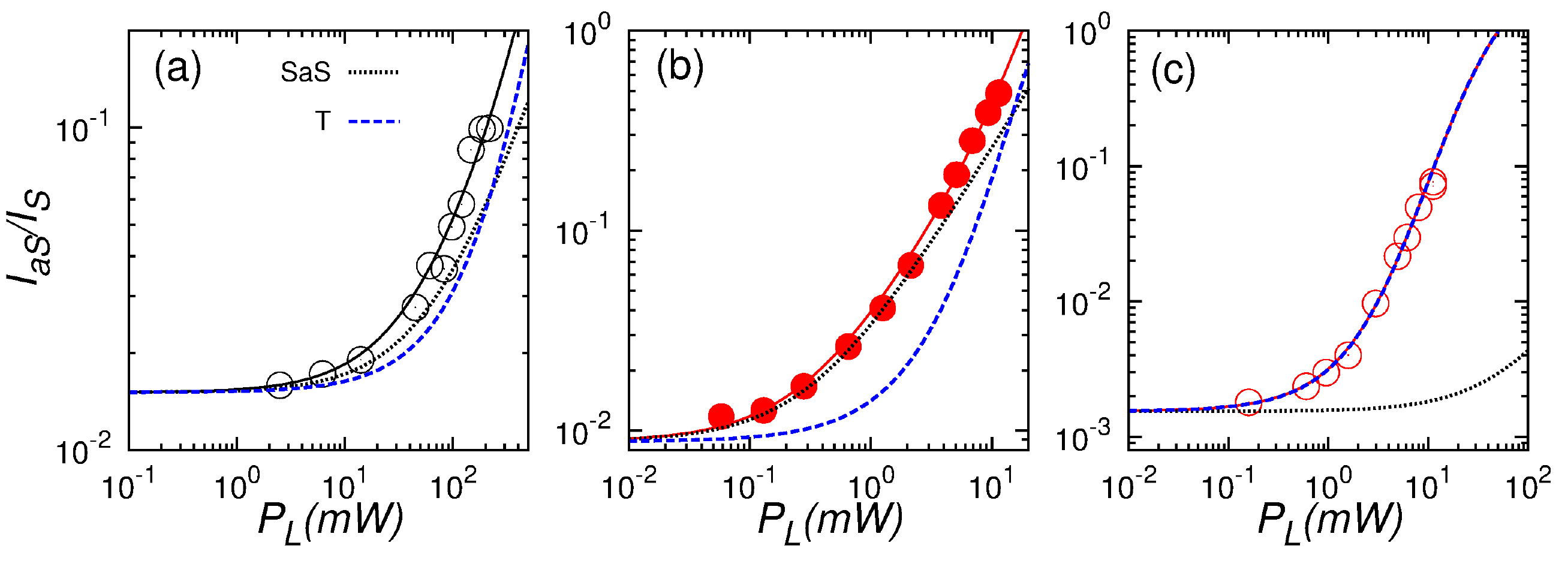}
\caption{\label{TvsSaS} Contribution from the thermal phonons and from the {\it SaS} phenomena in the fitting of $I_{aS}/I_{S}$ from (a) diamond, (b) tBLG and (c) AB-BLG. The symbols are the experimental data. The solid black lines are the best fits using Eq.(7) from the paper, with the parameters shown in Table I. The dashed blue and the dotted black lines give the separate contribution from the thermal phonons and from the {\it SaS} phenomena, respectively.}
\end{figure}

Figure\,\ref{TvsSaS} presents the separate contribution from the thermal phonons (dashed blue lines) and from the {\it SaS} phenomena (dotted black lines) to the $I_{aS}/I_{S}$ data fitting (solid black lines) for diamond (a), tBLG (b) and AB-BLG (c). As discussed in the paper, for diamond and tBLG, the {\it SaS} phenomena dominates until large laser powers, when the thermal phonons overcome. For AB-BLG, the {\it SaS} phenomena is minor.

It remains to be discussed the effect of $\lambda_S \neq \lambda_{aS}$, which is important when resonance effects are in place, i.e. when either Stokes or anti-Stokes has a stronger coupling to the
pumping field. Considering 
$\frac{\lambda_{aS}}{\lambda_{S}}=\varepsilon$
and setting $\lambda_S=\lambda$ and $\lambda_{aS}=\varepsilon \lambda$, once again as long as $\lambda \neq 0$, Eq.(7) in the paper changes to:
\begin{eqnarray}
\frac{\langle n_{aS}\rangle_{ss}}{\langle n_{S}\rangle_{ss}}=\varepsilon^2\left(\frac{2n_0\tilde\gamma(\tilde\gamma+1)-(1-2\tilde\gamma+n_0(3+\varepsilon^2))\tilde\lambda^2_{\alpha}}{2(n_0+1)\tilde\gamma(\tilde\gamma+1)-(1+n_0+(2+3n_0+2\tilde\gamma)\varepsilon^2)\tilde\lambda^2_{\alpha}}\right)\,.
\label{eqvar}
\end{eqnarray}
For $\tilde\gamma\gg \{n_0,1,\varepsilon^2\}$, this expression is well approximated by
\begin{eqnarray}
\frac{\langle n_{aS}\rangle_{ss}}{\langle n_{S}\rangle_{ss}}&=&\varepsilon^2\frac{n_0}{n_0+1}
\left(\frac{1+\frac{\tilde\lambda^2_{\alpha}}{\tilde\gamma n_0}}{1-\left(\frac{1}{2\tilde\gamma^2}+
\frac{\varepsilon^2}{\tilde\gamma (n_0+1)}\right)\tilde\lambda^2_{\alpha}}\right)\nonumber
\label{eqvar}
\end{eqnarray}
which approximately reduces to:
\begin{equation}
\frac{\langle n_{aS}\rangle_{ss}}{\langle n_{S}\rangle_{ss}}\approx\varepsilon^2\frac{n_0}{n_0+1}\left(1+\frac{1}{n_0}\frac{\tilde\lambda^2_{\alpha}}{\tilde\gamma}\right)\left(1+\frac{\varepsilon^2}{n_0+1}\frac{\tilde\lambda^2_{\alpha}}{\tilde\gamma}\right).
\label{eqvar2}
\end{equation}

\noindent Finally, the intensities ratio, in terms of $P_0=\mathcal{A}\gamma\gamma_c/2\lambda^2$ and when with $\varepsilon=\lambda_{aS}/\lambda_{S}<1$, is given by
\begin{eqnarray}\label{eq:013a}
\frac{I_{aS}}{I_{S}}&=&C'\varepsilon^2\frac{n_0}{n_0+1}\left(1+\left(\frac{1}{n_0}+\frac{\varepsilon^2}{(n_0+1)}\right)C_{SaS}P_L+\frac{\varepsilon^2}{n_0(n_0+1)}C_{SaS}^2P_L^2\right)\,
\label{eqvar}
\end{eqnarray}

\noindent with $C_{SaS}=1/2P_0$. It is straightforward to compute the other case, i.e., when $\lambda_{s}<\lambda_{aS}$. In that case
if we redefine $\varepsilon = \lambda_{s}/\lambda_{aS}<1$, the expression for the intensities ratio is:
\begin{eqnarray}\label{eq:013b}
\frac{I_{aS}}{I_{S}}&=&C'\varepsilon^{-2}\frac{n_0}{n_0+1}\left(1+\left(\frac{\varepsilon^2}{n_0}+\frac{1}{(n_0+1)}\right)C_{SaS}P_L+\frac{\varepsilon^2}{n_0(n_0+1)}C_{SaS}^2P_L^2\right).
\label{eqvar}
\end{eqnarray}

In non-resonant scattering, $\lambda_S = \lambda_{aS}$ is expected
due to time-reversal symmetry. This is the case for diamond \cite{Jorio-1}.
If resonance with electronic states exists, but there is not much difference between the resonance with
the {\it S} and {\it aS} photon emission, then $\lambda_S \sim \lambda_{aS}$. This is the case of AB-BLG, where the density of states
does not vary too much between the {\it S} and {\it aS} photon energies \cite{Jorio-2}.
However, in the case of tBLG, the electronic structure is engineered to exhibit specific resonance with the
{\it aS} photon emission \cite{Jorio-2}, i.e. $\lambda_{aS} > \lambda_{S}$. Therefore, strictly speaking, Eq.\,\ref{eq:13b} from this Supplemental
Information should be used to fit the data, rather than Eq.\,7 from the paper. The value of $\varepsilon$ is easily obtained from the
data if we consider that this parameter rules the enhancement of $C'$ from tBLG as compared to AB-BLG. From table I in the paper and
Eq.\,\ref{eq:13a} from this Supplemental Information we obtain $\varepsilon^2 = 3.5/20$. We now add $\varepsilon = 0.418$ in
Eq.\,\ref{eq:13a} from this Supplemental Information and fit the tBLG data with $C^{\prime} = 3.5$, which is consistently the value for AB-BLG.
This means that the effect of resonance in tBLG is now accounted for with $\varepsilon$.
The data fitting gives $(\lambda^2/\gamma\gamma_c){\cal A}^{-1}=(8.1\pm 0.5)\times 10^{-3}$\,mW$^{-1}$ and ($T{\rm [K]} = 295 + (15\pm 1)P_L{\rm [mW]}$). Therefore, considering the possibility of
$\lambda_S \neq \lambda_{aS}$ in our analysis actually enhances the importance of the {\it SaS} phenomena in tBLG by almost one order of magnitude (compare with the value for $(\lambda^2/\gamma\gamma_c){\cal A}^{-1}$ in Table I). Furthermore, the dependence on temperature decreases to half of the value obtained for AB-BLG, which is ($T{\rm [K]} = 295 + 30P_L{\rm [mW]}$), consistent with the G band frequency dependence observed in Fig.7 of Ref.\cite{Jorio-2}. Although this picture is fully consistent, we chose to
keep the simpler equation in the paper since the general picture does not change, and the discussion is simpler.

\end{document}